\documentclass[preprint,prb]{revtex4}

\begin{document}

\title{A density functional perspective for one-particle systems}
\author{H. L. Neal\cite{EMAIL}}

\affiliation{Department of Physics\\
and\\
Center for Theoretical Studies of Physical Systems\\
Clark Atlanta University\\
Atlanta, Georgia 30314}

\begin{abstract}
Density functional theory is discussed in the context of one-particle
systems. We show that the ground state density $\rho _0(x)$ and energy $E_0$
are simply related to a family of external potential energy functions with
ground state wave functions $\psi _n(x)\propto~\rho _0(x)^n$ and energies $%
E_n=2nE_0$ for certain integer values of $n$.
\end{abstract}

\maketitle

\section{INTRODUCTION}

The objective of density functional theory\cite{HK} (DFT) is to transform
the many-particle problem into an equivalent one-particle problem.
Therefore, it seems pedagogically reasonable to give an introductory level
presentation of DFT, in the context of one-particle systems, that (briefly)
ovoids references to the complications imposed by systems with two or more
particles. Several papers\cite{HLN} devoted to DFT have appeared in this
journal.

The basic idea of DFT is to transform from the ground state wave function $%
\psi _0$ to the one-particle density function $\rho _0$ as a focus of
attention. This has obvious advantages for many-particle systems. For
one-particle systems, it allows the exact ground state energy $E_0$ to be
determined via the variational principle with respect to the one-particle
density. For one-particle systems 
\begin{equation}
\rho _0(x)=\left| \psi _0(x)\right| ^2,  \label{RHO}
\end{equation}
where the density function satisfies the normalization condition 
\begin{equation}
\int \rho _0(x)dx=1.  \label{RHO2}
\end{equation}
The wave function $\psi _0$ satisfies the time independent Schr\"{o}dinger
equation 
\begin{equation}
\left[ -\frac{\hbar ^2}{2m}\frac{d^2}{dx^2}+V(x)\right] \psi _0(x)=E_0\psi
_0(x),  \label{SE}
\end{equation}
where $V(x)$ is the external potential energy function for the particle of
mass $m$. We consider systems where $\psi _0$ is real, so that the energy
may be written as 
\begin{equation}
E_0=\int \sqrt{\rho _0(x)}\left[ -\frac{\hbar ^2}{2m}\frac{d^2}{dx^2}%
+V(x)\right] \sqrt{\rho _0(x)}dx.
\end{equation}
We now introduce the energy density functional 
\begin{equation}
E[\rho ]=T[\rho ]+\int \rho (x)V(x)dx,
\end{equation}
where the kinetic energy density functional $T[\rho ]$ is 
\begin{equation}
T[\rho ]=\int \sqrt{\rho (x)}\left( -\frac{\hbar ^2}{2m}\frac{d^2}{dx^2}%
\right) \sqrt{\rho (x)}dx.
\end{equation}
The arbitrary density function $\rho $ satisfies the same normalization
condition as $\rho _0$ in Eq.~(\ref{RHO2}). According to the Hohenberg-Kohn%
\cite{HK} (HK) theorem, (1) \emph{the external potential} $V(x)$ \emph{is
determined uniquely}\cite{HK2}\emph{\ by} $\rho _0(x)$, and (2) \emph{the
energy density functional} $E[\rho ]$ \emph{satisfies the condition} 
\begin{equation}
E[\rho ]\ge E_0.  \label{DFT2}
\end{equation}
Variationally speaking, the minimum value of $E[\rho ]$ with respect to $%
\rho $ yields the ground state energy $E_0=$ $E[\rho _0]$. The density
functional $T[\rho ]$ may be written more conveniently as 
\begin{equation}
T[\rho ]=\frac{\hbar ^2}{8m}\int \left( \frac 1{\rho (x)}\frac{d\rho (x)}{dx}%
\right) ^2\rho (x)dx
\end{equation}
According to the variational principle 
\begin{equation}
\delta \left( E[\rho ]-\lambda \int \rho (x)dx\right) =0,
\end{equation}
where $\lambda $ is a Lagrange multiplier. In terms of variational (or
functional) derivatives\cite{ARF} 
\begin{equation}
\frac{\delta E[\rho ]}{\delta \rho (x)}=\lambda ,
\end{equation}
so that the equation for the density becomes 
\begin{equation}
-\frac{\hbar ^2}{8m}\left\{ \frac{d^2}{dx^2}\ln \rho (x)^2+\left( \frac
d{dx}\ln \rho (x)\right) ^2\right\} +V(x)=\lambda .  \label{R1}
\end{equation}
This equation may be rewritten in a form similar to Eq.~(\ref{SE}): 
\begin{equation}
\left[ -\frac{\hbar ^2}{2m}\frac{d^2}{dx^2}+V_{\text{eff}}(x;[\rho ])\right]
\rho (x)=2\lambda \rho (x),  \label{NLSE}
\end{equation}
where the effective potential $V_{\text{eff}}$ is 
\begin{equation}
V_{\text{eff}}(x;[\rho ])=2V(x)+\frac{\hbar ^2}{4m}\left( \frac 1{\rho
(x)}\frac d{dx}\rho (x)\right) ^2.  \label{VEFF}
\end{equation}
Equation (\ref{NLSE}) is an example of the nonlinear Schr\"{o}dinger
equation that is encountered in DFT calculations. The standard method of
solution is to solve it iteratively, starting with an initial guess for $V_{%
\text{eff}}$. At the solution point $\rho =\rho _0$ and $\lambda =E_0$.
Equation (\ref{R1}) may be transformed into a first order differential
equation for $y(x)=d\ln \rho (x)/dx$: 
\begin{equation}
-\frac{\hbar ^2}{8m}\left\{ 2\frac d{dx}y(x)+y(x)^2\right\} +V(x)=\lambda .
\label{R2}
\end{equation}
This is a version of the Riccati\cite{RAC1} equation. A solution method for
this equation has been discussed in some detail by Haley.\cite{HALEY}

In Sec.~\ref{SEC2}, we derive equations that give the external potentials
and ground state energies for the family of ground state wave functions $%
\psi _n(x)\propto \rho _0(x)^n$, where $n=2^j$ and $j$ ($j\ge 0$) is an
integer. In Sec.~\ref{SEC3}, we apply the results of Sec.~\ref{SEC2} to
several well known systems. In Sec.~\ref{SEC4}, we give a summary and some
closing remarks.

\section{GENERATING EXTERNAL POTENTIALS WITH THE GROUND STATE DENSITY}

\label{SEC2}

In this section we demonstrate one consequence of the HK theorem: The ground
state density $\rho _0(x)$ determines a family of external potentials and
their respective ground state energies. We give some examples in the next
section.

Consider the wave functions given by 
\begin{equation}
\psi _{n}(x)=c_{n}\rho _{0}(x)^{n},  \label{PHIN}
\end{equation}
where the integer $n=2^{j}$ for the integer $j$, and the constants $c_{n}$
are chosen to normalize $\psi _{n}(x)$. Each ground state wave function
given by Eq.~(\ref{PHIN}) satisfies the Schr\"{o}dinger equation 
\begin{equation}
\left[ -\frac{\hbar ^{2}}{2m}\frac{d^{2}}{dx^{2}}+V_{n}(x)\right] \psi
_{n}(x)=E_{n}\psi _{n}(y),  \label{SE2}
\end{equation}
where $E_{n}$ is the ground state energy for the external potential $V_{n}(x)
$. For example, applying Eqs. (\ref{NLSE}) and~(\ref{VEFF}) iteratively for $%
n=1$, $2$, and $4$ yields 
\begin{eqnarray}
V_{1}(x) &=&V_{\text{eff}}(x;[\rho _{0}])  \nonumber \\
&=&2V(x)+\frac{\hbar ^{2}}{4m}\left( \frac{1}{\rho _{0}(x)}\frac{d}{dx}\rho
_{0}(x)\right) ^{2};  \label{V1} \\
V_{2}(x) &=&2V_{1}(x)+\left( 2\right) ^{2}\frac{\hbar ^{2}}{4m}\left( \frac{1%
}{\rho _{0}(x)}\frac{d}{dx}\rho _{0}(x)\right) ^{2}; \\
V_{4}(x) &=&2V_{2}(x)+\left( 4\right) ^{2}\frac{\hbar ^{2}}{4m}\left( \frac{1%
}{\rho _{0}(x)}\frac{d}{dx}\rho _{0}(x)\right) ^{2}.
\end{eqnarray}
In the Appendix, we show that the family of external potential energy
functions and ground state energies are given, respectively, by 
\begin{eqnarray}
V_{n}(x) &=&2nV(x)+\left( 2n-1\right) n\frac{\hbar ^{2}}{4m}\left( \frac{1}{%
\rho _{0}(x)}\frac{d}{dx}\rho _{0}(x)\right) ^{2};  \label{VN} \\
E_{n} &=&2nE_{0}.  \label{EN}
\end{eqnarray}

\section{EXAMPLES}

\label{SEC3}

We have applied Eq.~(\ref{VN}) to (\emph{i}) the particle in a box, (\emph{ii%
}) the harmonic oscillator, and (\emph{iii}) the attractive delta potential.
The results for these systems are summarized in Table \ref{TAB1}.

\section{SUMMARY AND REMARKS}

\label{SEC4}

We have presented DFT for one-particle systems. As a consequence of the HK
theorem, the ground state density $\rho _0(x)$ determines the external
potential $V(x)$ and a family of related external potentials $V_n(x)$ given
by Eq.~(\ref{VN}). Moreover, the ground state energies for these external
potentials are given by Eq.~(\ref{EN}). These results results are directly
applicable to higher dimensional system by replacing $x$ by the position
vector $\mathbf{r}$ and the derivative operator $d/dx$ by the vector
operator $\nabla $.

We hope that our presentation will be useful and insightful to teachers and
students of quantum mechanics.

\begin{acknowledgments}
The author thanks Professor R. E. Mickens
for helpful discussions.
\end{acknowledgments}

\appendix

\section{A DERIVATION OF EQUATIONS (\ref{VN}) AND (\ref{EN})}

\subsection{Equation (\ref{VN})}

The equations for the external potentials discussed in Sec.~\ref{SEC2} may
be transformed into the linear difference equation 
\begin{equation}
z_{j+1}=2z_j+\left( 2^{j+1}\right) ^2f(x),  \label{DIFE}
\end{equation}
where $z_j=V_{2^j}(x)$ and the function $f(x)$ is 
\begin{equation}
f(x)=\frac{\hbar ^2}{4m}\left( \frac 1{\rho _0(x)}\frac d{dx}\rho
_0(x)\right) ^2.
\end{equation}
The solution\cite{LT} to Eq.~(\ref{DIFE}) is 
\begin{eqnarray}
z_j &=&2^jz_0+2^jf(x)\sum_{i=0}^{j-1}2^{i+1}  \nonumber \\
&=&2^jz_0+2^{j+1}f(x)\left( 2^j-1\right) .
\end{eqnarray}
In terms of $n=2^j$%
\begin{equation}
V_n(x)=nV_1(x)+2n\left( n-1\right) f(x).
\end{equation}
Inserting the expression for $V_1(x)$ given by Eq.~(\ref{V1}) into this
equation yields Eq.~(\ref{VN}).

\subsection{Equation (\ref{EN})}

Consider the two equations resulting from the insertion of the ground state
densities $\rho _0(x)$ and $\rho _n(x)\propto \rho _0(x)^{2n}$,
respectively, into Eq.~(\ref{R1}): 
\begin{eqnarray}
-\frac{\hbar ^2}{8m}\left\{ \frac{d^2}{dx^2}\ln \rho _0(x)^2+\left( \frac
d{dx}\ln \rho _0(x)\right) ^2\right\} +V(x) &=&E_0; \\
-\frac{\hbar ^2}{8m}\left\{ 2n\frac{d^2}{dx^2}\ln \rho _0(x)^2+4n^2\left(
\frac d{dx}\ln \rho _0(x)\right) ^2\right\} +V_n(x) &=&E_n.
\end{eqnarray}
Multiplying the first equation by $2n$ and then subtracting the resulting
equation from the second equation yields Eq.~(\ref{EN}).

\pagebreak

\begin{center}
\textbf{TABLES}
\end{center}

\begin{table}[h]
\centering
\caption{The external potential $V(x)$, the ground state density $\rho _0 (x)$,
the ground state energy $E _0$, and the external potential $V_n (x)$ 
given by Eq.~(\ref{VN}) for (\emph{i}) the particle
in a box, (\emph{ii}) the harmonic oscillator, and
(\emph{iii}) the attractive delta
potential.} \label{TAB1} 
\begin{tabular}{|c|c|c|c|c|}
\hline\hline
\textbf{System} & $V(x)$ & $\rho _0(x)$ & $E_0$ & $V_n(x)$ \\ \hline
(\emph{i}) & $\left\{ 
\begin{tabular}{c}
$0$ for $\left| x\right| <L/2$ \\ 
$\infty $ otherwise
\end{tabular}
\right. $ & $\left( 2/L\right) \cos ^2\left( \pi x/L\right) $ & $\left(
\hbar \pi /L\right) ^2/2m$ & $2n\left( 2n-1\right) E_0\tan ^2\left( \pi
x/L\right) $ \\ \hline
(\emph{ii}) & $m\omega ^2x^2/2$ & $\sqrt{m\omega /\pi \hbar }\exp \left(
-m\omega x^2/\hbar \right) $ & $\hbar \omega /2$ & $4n^2V(x)$ \\ \hline
(\emph{iii}) & $-g\delta (x)$ & $\left( mg/\hbar ^2\right) \exp \left(
-2mg\left| x\right| /\hbar ^2\right) $ & $-mg^2/2\hbar ^2$ & $%
2nV(x)-2n\left( 2n-1\right) E_0$ \\ \hline\hline
\end{tabular}
\end{table}


\begin{thebibliography}{a)}
\bibitem{HK2}  If the potential $V(x)$ is determined by $\rho _0(x)$, then
the potential $V^{\prime }(x)=V(x)+C$ is also determined by $\rho _0(x)$,
where $C$ is a constant.
\end{thebibliography}
\end{document}